# LOOKING BEYOND CONTENT:
# SKILL DEVELOPMENT FOR ENGINEERS

EDWARD F. REDISH

*Department of Physics*
*University of Maryland*

KARL A. SMITH

*Department of Engineering Education*
*Purdue University*

Current concerns over reforming engineering education have focused attention on helping students develop skills and an adaptive expertise. Phenomenological guidelines for instruction along these lines can be understood as arising out of an emerging theory of thinking and learning built on results in the neural, cognitive, and behavioral sciences. We outline this framework and consider some of its implications for one example: developing a more detailed understanding of the specific skill of using mathematics in modeling physical situations. This approach provides theoretical underpinnings for some best-practice instructional methods designed to help students develop this skill and provides guidance for further research in the area.

## 1. Introduction

The increasing importance of technology in our modern economic system and the increased globalization of scientific and technological ideas, development, and production have focused national attention on the education of scientists and engineers. This comes at a time when research into thinking and learning has produced both dramatic new understandings and productive new technologies for building effective teaching and learning environments. The synergy of bringing together basic research on cognition with education research focused directly on the disciplines of science and engineering promises continued improvement of our instructional effectiveness in the near future.

In this article, we focus on preparation of students for professional practice. Education research into learning at the tertiary level is becoming increasingly widespread. This is particularly the case in physics where research on physics teaching and learning has been a growing component of physics departments for nearly three decades. As of this writing, there are more than 35 physics departments that have research groups studying the teaching and learning of its discipline and hundreds more use learning environments developed by these researchers. According to the National Academy of Engineering CASEE website, there are five U.S. schools with PhD programs in engineering education and more than 20 schools with engineering education centers, and interest in engineering education research appears to be growing.

In this article, we consider the goals for engineering education that arise out of the changing workplace environment and analyze them in the light of an emerging theoretical framework for modeling thinking and learning that is based on recent developments in cognitive science. This theoretical framework permits the development of a model of cognitive mechanism that links broad goals and guidelines with specific instructional models. We apply this framework to the example of one important cognitive skill for engineering students: mathematical modeling.

## 2. What Engineering Students Need to Learn Today

Many engineering undergraduate programs today look remarkably similar to those described in the 1918 Mann Report, *A Study of engineering education* (Mann, 1918). Current concerns about engineering programs – overcrowded curricula, need for integration of theory and practice, need for better student retention, difficulty in assessment, professional development of faculty – are similar to those expressed nearly a century ago. Numerous studies and reports now call for a different type of engineering graduate. Thoughtful studies abound, including:

- A 1997 Engineering Futures Conference co-sponsored by Boeing and Rensselaer Institute, which resulted in a provocative set of characteristics of a global engineer (Boeing, 1997).
- The National Academy of Engineering Engineer of 2020 report (National Academy of Engineering, 2004) and Educating the engineer of 2020 (National Academy of Engineering, 2005).
- The National Engineering Education Research Colloquies (Steering Committee of The National Engineering Education Research Colloquies, 2006)
- Commissioned papers for the National Center on Education and the Economy, including, "Rethinking and Redesigning Curriculum, Instruction and Assessment" (Pellegrino, 2006).

These studies are converging on a view of engineering education that not only requires students to develop a grasp of traditional engineering fundamentals, such as mechanics, dynamics, mathematics, and technology, but to also develop the skills associated with learning to imbed this knowledge in real-world situations. This not only demands skills of creativity, teamwork, and design, but in global collaboration, communication, management, economics, and ethics. Furthermore, the rapid pace of change of technology seems fated to continue for many decades to come. This will require the engineers we are training today to learn to be lifelong learners and to learn to develop (in the felicitous phrase (Hatano and Inagaki, 1986; Pellegrino, 2006) *adaptive expertise*.

One model that attempts to identify elements that need to be included in a modern curriculum in order to reach the desirable outcomes identified in *The Engineer of 2020* (National Academy of Engineering, 2004) is the "Three Curricular Pillars" developed at Purdue University for guiding their curricular reforms, shown in Figure 1 (Jamieson, 2007). This, and other reform plans like it, illustrate the complexity of the educational issues that need to be considered today.

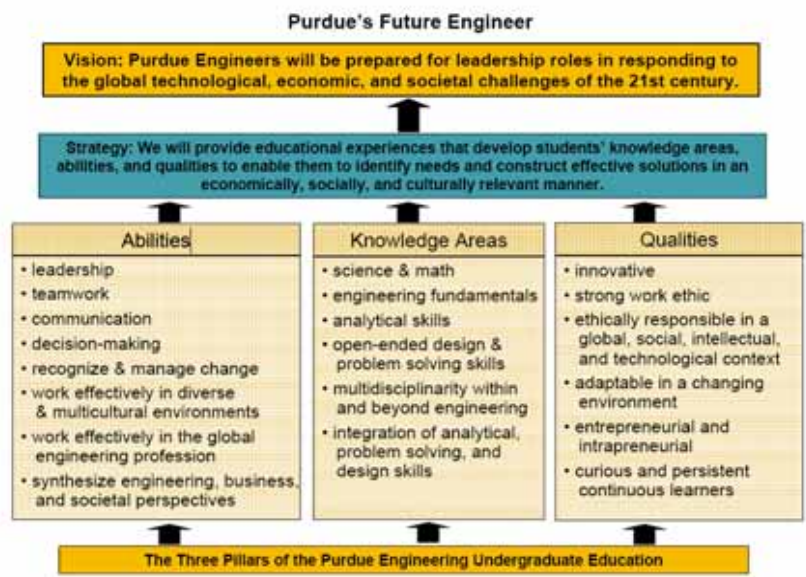

Figure 1. The Purdue Pillars of Engineering Undergraduate Education.

The engineering education community has been highly proactive in conducting studies of engineering education and proposing goals. The community has been less effective in figuring out how to achieve these goals and in changing current practice. Our thesis is that engineering education research as well as research from related disciplines – physics, chemistry, medicine, nursing – can provide the theory and body of knowledge needed to guide the community to achieving its goals. Pellegrino stresses that three educational components – curriculum, instruction, and assessment – not only have to be developed to support these new goals, they have to be aligned in support of each other. He states (Pellegrino, 2006, p. 3).

> *"Alignment is difficult to achieve, however. Often what is lacking is a central theory about the nature of learning and knowing in a given domain of knowledge and expertise around which the three functions can be coordinated.*
>
> *Most current approaches to curriculum, instruction, and assessment are based on theories and models that have not kept pace with modern knowledge of how people learn. They have been designed on the basis of implicit and highly limited conceptions of learning. Those conceptions tend to be fragmented, outdated, and poorly delineated for domains of subject matter knowledge. Alignment among curriculum, instruction, and assessment could be better achieved if all three are derived from a scientifically credible and shared knowledge base about cognition and learning in the subject matter domains."*

In the next section, we outline the beginnings of a theoretical framework, based on modern studies of human thinking and learning gleaned from the converging knowledge of the neural, cognitive, and behavioral sciences.

## 3. What Education Research Teaches Us

### A. *How People Learn*

Much has been learned over the past few decades about how people learn and how this plays into science education. Although there is an extensive literature (see Bransford, Brown, Cocking, Donovan and Pellegrino, 2000 for an overview) Pellegrino summarizes the main threads into three broad results (Pellegrino, 2006, 3-5).

1. *Constructivism – Students come to the classroom with preconceptions about how the world works which include beliefs and prior knowledge acquired through various experiences.*
2. *Knowledge organization – To develop competence in an area of inquiry, students must: (a) have a deep foundation of factual knowledge, (b) understand facts and ideas in the context of a conceptual framework, and (c) organize knowledge in ways that facilitate retrieval and application.*
3. *Thinking about thinking – A "metacognitive" approach to instruction can help students learn to take control of their own learning by defining learning goals and monitoring their progress in achieving them.*

These principles, while largely based on phenomenological observations in the classroom, are consistent with the emerging theoretical framework of the dynamics of thinking and learning. Moreover, interpreting these ideas through the framework refines them and gives additional power for understanding how they arise, how they play out in particular situations, and how we can create more effective learning environments. For a summary and extensive specific examples in physics, see McDermott and Redish (1999), Knight (2002), Thacker (2003), Redish (2003), and Redish and Cooney (2007).

These results and the studies that led to them tend to focus on what we might call "the learning of content" – helping students to understand the basic concepts and meaning associated with the "table of contents" of the class. This is extremely important. (See Streveler, Litzinger, Miller, and Steif, 2008). But every science and engineering instructor knows (and as Pellegrino's second and third principles stress) a good knowledge of the facts, equations, and even concepts is only the beginning. What matters more is that the students learn the *practice* of science and engineering – not only the knowledge needed but how to use that knowledge in authentic contexts. Much of the interest in engineering education today is on understanding and developing these skills of practice.

Often what our students learn in our classes about the practice of science and engineering is implicit and may not be what we want them to learn. For example, a student in an introductory engineering physics class may learn that memorizing equations is important but that it is not important to learn the derivation of those equations or the conditions under which those equations are valid. This meta-message may be sent unintentionally. The student may, for example, read this conclusion from the fact that "tricky" problems, in which it looks like an equation should work but it does not in fact apply, will never appear on an exam. Or a student may learn that any numbers they need will always be given to them and that the quantification of their personal experience is unnecessary and irrelevant. This may have the result that they learn not to test their calculated results for plausibility based on experience, something few engineering physics teachers intend students to learn from an introductory class. Attitude and expectation surveys have shown both of the above misunderstandings increase as a result of traditional instruction in introductory physics for engineers (Redish, Saul and Steinberg, 1998).

Many of the items discussed in the studies cited in section II and figure 1 that lead to adaptive expertise are *process skills* – knowledge of how to employ factual knowledge in practice. Others are what we might call *epistemological skills* – concerned with issues of how do you know the validity of the knowledge you have and how do you construct new knowledge in situations in which you don't immediately know the answer (problem solving). How these active skills and their development relate to more passive factual knowledge rests on our understanding of how the mind works and handles its knowledge in situations of practice.

## 4. Modeling Thinking and Learning

The phenomenological research observations and attempts to create and evaluate effective teaching environments in science and engineering (Duit, 2007) are the "experimental" and "engineering" aspects of a science of science education. Yet we know well, that although engineers can make much progress phenomenologically, more power and understanding develops when one understands the underlying mechanisms. Such an understanding is provided by theory. In education, an empirically based theory of thinking and learning has so far played little role. But the understanding of the mechanisms of thinking and learning has grown dramatically in the past few decades and one may hope to begin the development of a theoretical framework that can guide, constrain, and learn from our experiments (Redish, 2005). Such a theoretical frame allows us to bridge the different conceptual models discussed in the literature (see Streveler et al., 2008) and to see whether the students might be responding by activating a "theory" or a "facet" as an empirical question to be determined by observation.

Thinking and learning are immensely complex processes. People may know something but not be able to access it. Thinking is dynamic, with bits of knowledge freely associating and popping up in ways that can depend strongly on the individual's perception of the context and environment. Developments in neuroscience, cognitive science, behavioral science, and education have begun to provide us with improved ways of talking about the way people think and learn. Four broad concepts can help us talk about the identification, teaching, and evaluation of skills in a significant way.

- Activation
- Association
- Compilation
- Control

### A. *Activation*

According to currently accepted models in neuroscience and cognitive science, essentially everything individuals have learned are stored in the state of their system of neurons, in particular, in the strength of synaptic connections (Kandel, Schwartz and Jessell, 2000). A "thought" or perception of a particular object corresponds to the coordinated electrical activation of a set of neurons (propagation of periodic sharp voltage pulses called action potentials through wire-like extensions of neurons). The neural system is highly interconnected, with each neuron receiving up to thousands of inputs from other neurons through transducing connections called synapses. Whether a neuron fires at a particular time depends on the weighted sum of the input signals from other neurons. When a signal to fire comes from a pre-synaptic neuron and the post-synaptic neuron actually fires, the synaptic connection strengthens and becomes more effective – Hebb's rule (Hebb, 1949). Thus, clusters of neurons become entrained and tend to activate together in a coordinated (and synchronized) way (Fuster, 1999).

Such clusters may be treated as a unit when considered from the point of view of the perception of the thinker and of functionality. We refer to such a bit of knowledge as a *resource* (Hammer, 2000). The "facets" discussed in Steveler et al. (2008) are one kind of resource, but resources are a more general structure, for example, they could be epistemological rather than conceptual.

Hebb's rule provides a fundamental mechanism for learning. Much is now known about how strengthened synapses form (Kopec and Manilow, 2006). Usually, this implies that learning needs significant repetition over extended time periods. (Though the association structures discussed below can facilitate this and reduce learning times.)

The main features of our memory system can be thought about in terms of two broad structures: long-term memory (LTM) and working memory (WM) (Baddely, 1998). LTM is a vast data store that contains the elements of all of our knowledge and memories. However, it has a number of characteristics that make it less than ideal. (1) LTM can be highly compressed. Blocks of knowledge appear to be stored as guidelines, being reconstructed as needed with "standard" or "plug-in" elements (Kotre, 1996; McCelland, McNaughton and O'Reilly, 1995). This can lead to the conflating of distinct memories or the creation of false memory. (2) LTM does not have an effective search routine. Bringing items in LTM to consciousness (or pre-consciousness) requires activation of the neurons involved. This has to be done through chains of associations and may take long times – seconds, minutes, or more.

Working memory (WM) is the term used to describe the active mental manipulation of knowledge. WM draws on the knowledge stored in LTM (and indeed may just be the subset of LTM activated at a particular instant) but it is limited in two ways.

1. WM can only attend to and manipulate a small number of distinct elements at once. These elements are referred to as "chunks" and tend to be perceived by the thinker as distinct (though not necessarily irreducible) elements. These may be the traditional "7 ± 2" (Miller, 1956) or even fewer.
2. WM is labile and lasts only for seconds if not actively refreshed. (Think of trying to remember a telephone number while looking for a pencil.) This is in contrast to LTM, which can store information reasonably reliably for decades without intentional intervention.

The structures of working and long-term memory and the mechanism of reconstructive recall relate directly to Pellegrino's first principle. Without understanding the mechanism of constructivism, we might be tempted to look at students' preconceptions as incorrect, rigid, unary structures that must be purged before students can learn the "correct" results. A more detailed analysis, however, often shows that student errors, while being robust and predictable, are not as rigid as often thought.

For example, in introductory physics classes, students are often confused by Newton's third law. When a mosquito hits a car's windshield, how can the mosquito possibly exert as much force on the car as the car exerts on the mosquito! The mosquito is squashed and the car hardly feels it! The intuitions on which these conclusions are based are, of course, correct. But if the conclusions the students draw are seen as generated on the spot from combining knowledge that is easy to hand and naturally associated, the instructional response can be dramatically different and more effective (Elby, 2001). This is discussed in (Streveler et al., 2008) and we consider it again after we consider the issue of association.

## B. *Association*

The implications of the structure of neurons for thinking rely heavily on how clusters of neurons associate and activate together. When some environmental data is presented to the individual (a lesson, demonstration, problem), links between different resources may lead to chains of excitation of additional resources known as *spreading activation*. Depending on the strength of various links, the information and knowledge that is retrieved by this process may be appropriate or inappropriate.

Items in LTM may be difficult to access in appropriate circumstances if the relevant links and organizations have not been made. As a result, it is critically important to not only know what knowl-edge a student has in his or her head, but how that knowledge is structured and connected to other knowledge. One of the oldest and most commonly discussed structures used in cognitive science is the *schema* – a skeletal representation of knowledge abstracted from experience that organizes and guides the creation of particular representations in specific contexts (Bartlett, 1932; Rumelhart, 1981). A variety of specific structures have been proposed including coordination classes (diSessa and Sherin, 1998), local coherences (Sabella and Redish, 2007), and epistemic games (Tuminaro and Redish, 2007), to mention a few.

Associational paths are the key element underlying Pellegrino's second principle, knowledge organization. Varying the context can have a dramatic effect on what knowledge is activated. Combining this idea with the idea of activation can both explain student responses and provide hints towards constructing more effective instruction.

For example, in Elby's method of teaching Newton's 3[rd] law (Elby, 2001; Elby et al., 2007, students are asked the following pair of questions:

> A. Consider a heavy truck ramming into a parked, unoccupied car. According to *common sense*, which force (if either) is larger during the collision: the force exerted by the truck on the car, or the force exerted by the car on the truck? Explain the intuitive reasoning.

> B. Suppose the truck's mass is 2000 kg while the car's mass is 1000 kg. And suppose the truck slows down by 5 m/s during the collision. Intuitively, how much speed does the car gain during the collision? (Apply the intuition that the car reacts more during the collision, keeping in mind that the truck is twice as heavy.) Explain your intuitive reasoning" (Elby et al., 2007)

In the context of the first question, students tend to activate the resource "larger (and more active) means more effect" and conclude that the truck exerts a larger force. In the context of the second question, the student tends to activate the resource "compensation" and intuit that the momentum is conserved. When these are linked in the context of Newton's second law (rate of change of velocity equals net force divided by mass), it allows students to reinterpret their first answer and see the (previously ignored but crucial) role played by the difference in the object's masses.

When a one hour lesson (tutorial) using this approach replaces a standard recitation section, the students not only dramatically improve their results on the Newton 3 questions of the Force Concept Inventory (a conceptual survey of basic concepts in mechanics (Hestenes, Wells and Swackhamer, 1992) compared to traditional (from about 45% to 85%) but the fraction of the students giving the correct answer who stated that the answer "was intuitive" increased dramatically (from about 40% to about 88%). The result was that in the traditional class only ~25% of the students gave the correct answer and reported that it was intuitive compared to ~75% in the reformed class.

Analyzing thinking in terms of associational patterns leads us to suggest that when misconceptions are treated as unary and instruction focuses on trying to replace those misconceptions by correct ones, students build up alternative associational paths; one set of knowledge is activated specifically for a physics class but the other intuitive knowledge is not erased but remains for activation in all other situations. This is consistent with well-established results in cognitive science on conditioning, attempts to eliminate conditioning (extinction), and its reemergence (relapse) (Bouton, 2002).

## C. *Compilation*

One implication of Hebb's rule and the mechanism for learning is that items that are originally weakly associated may become very strongly tied together. We refer to this as *compilation* by analogy with the creation of a machine language executable program on a computer from higher-level code. Just as a programmer who is in possession of executables written by someone else may not be able to figure out exactly how an executable does what it does, a thinker with a compiled bit of knowledge may be unable to "open it up" and see its parts. For example, a competent reader of English finds it extremely difficult to look at the pattern of lines "yellow" and not read the word or associate it with the color (Stroop, 1935). Similarly, an expert engineer is unlikely to be able to look at a graph and not know immediately where the points of zero derivative lie. On the other hand, novice readers or students learning calculus may have to activate multiple bits of knowledge to construct the results that an expert does in a unary fashion without conscious thought.

The compilation of knowledge into tight irreducible parts yields new resources – chunks that can be used and easily manipulated in WM. Learners who have not yet compiled their knowledge on a particular topic may find the *cognitive load* of manipulating it exhausts a significantly larger fraction of WM than an expert might think. It is often difficult for thinkers who have compiled knowledge to appreciate what they have bound together in their own thoughts. This means that the instructor who is an expert in a subject must learn to *reverse engineer* their knowledge – to use a conscious analysis in order to make sense of what a student will need to go through to learn a subject.

Such reverse engineering (with careful input from observing the actual functioning of students) can lead to surprises. In one example (Redish, Scherr and Tuminaro, 2006), we observed a group of students solving the simple physics problem shown in figure 2. A physics instructor can often solve this in a beat without thinking. The students we observed took 45 minutes to solve it.

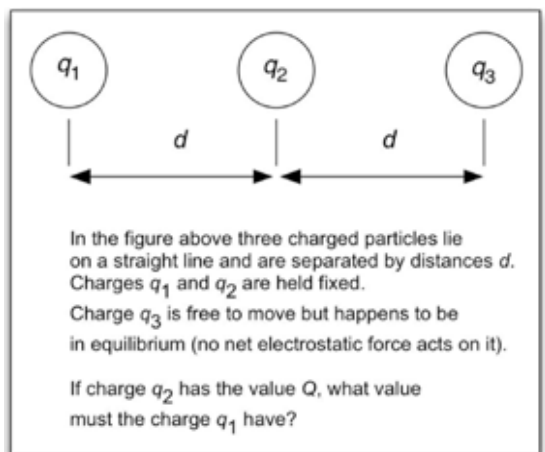

Figure 2. A simple physics problem.

Observing the video, it was clear that much of the work the students were doing was not directed at the problem itself but at recalling and working on compiling the many conceptual ideas that are required to make sense of the problem – a Newtonian analysis of forces (Which of the forces in the problem do we have to consider and how do they combine?), raising questions about the character of electric fields (Does having a charge in the middle block the effect of the charge on the left on the charge on the right?), etc. The Coulomb law of electric forces itself was not a problem. All the students knew it and could write it down. Viewing this work helped us reverse engineer the critical knowledge that we as experts had compiled and were essentially unaware of using.

Understanding the process of compilation is an essential element for an instructor trying to apply Pellegrino's first two principles – constructivism and knowledge organization.

### D. *Control*

The fourth broad principle we draw from the sciences of human thinking is *control*. This is the critical element in understanding the mechanisms underlying metacognition – and it helps us to understand that not all of the relevant mechanisms that we might classify with "thinking about thinking" are under conscious control. Some fall under the rubric of "expectations." These can lead students to explicitly fail to use appropriate knowledge that they indeed possess.

The neural system is characterized by *feedforward* and *feedback* processes (Kandel, Schwartz and Jessell, 2000). In the first, information brought into the system is sent on to processing elements and transformed for purposes of interpretation and activity. But feedback occurs everywhere and at all levels. Partially processed data are sent forward over high speed lines to places where partial information can be processed, compared against knowledge in LTM, and the results sent back to affect and modify the processing that is being carried out on the data stream. Such complex feedback structures are referred to in the neural and cognitive science literatures as *executive control* (Miyake and Shah, 1999). (This should not be read to imply the existence of a "guiding homunculus" in the brain.)

The implication of this at a higher level is that we never bring to bear everything we know about a situation. The external world has too much going on and we have too much data flowing in to be able to pay attention to all the inputs we receive given our limited attentional and processing resources. Over our years of experience with the physical and social world, we develop schemas that guide, select, and suppress what we notice and attend to. Preliminary data activate (typically in a tacit, unconscious way) these schemas and they determine where to put increased attention, which bits of our LTM and skills to activate and bring to bear, and what to suppress. These control schemas have three important consequences: they create context dependence, they give us a variety of resources for building new knowledge and solving problems, and they control which of these resources we bring to bear in given circumstances. We will see that these tell us that understanding the mechanisms involved in Pellegrino's third principle is much more subtle and complex than one might at first guess.

### 1) How context dependence is determined: Schemas

The schemas we have learned throughout our lives allow us to quickly and easily activate the responses appropriate for a particular situation. They code and guide the implementation of our expectations. When we walk into a restaurant, clues available from a quick scan can govern a broad set of our responses, even if we have never been in that restaurant before. A menu on the wall, a counter with the cooks behind it and a clerk ready to take your order and your cash will activate your "fast food" responses, including a willingness to eat with your hands, the knowledge that you will have to pay before you eat, that you may eat in our take your food out, etc. On the other hand, a *maitre de* carrying menus and leading you to a table with a tablecloth and an elegant place setting will activate your "fancy restaurant" responses, including the knowledge that you may *not* eat most things with your hands, that you will pay after you eat, that you are expected to eat at least a part of your meal sitting at the table (though you may request a bag or a box to take uneaten food with you), etc. It may even affect your posture and the pace of your walk (Bartlett, 1932; Scherr, in press).

Expectation schemas can have a big impact in how a student responds in a classroom. During the past year, one of us (EFR) visited a number of universities while on sabbatical and viewed a number of innovative classrooms. In one upper division physics classroom, the instructor divided the class into groups of 3-4 and gave them an estimation problem to work on. The problem involved estimating the amount of gasoline needed to carry a car over a particular route using energy conservation. The problem included the statement "a C-C bond has an energy of ~1 eV." In one group of students, one of them said, "Oh! I know the chemical formula for octane, so with that number we can count the number of carbon bonds broken in the burning and estimate how much energy is released in burning one gallon of gasoline." He proceeded to attack the calculation with enthusiasm. In a second group, one of the students commented, "Carbon bond? This is a physics class. They can't expect us to know chemistry. Let's just take 1 eV as the amount of energy given up per molecule." These two approaches reflect two different sets of expectations for what are appropriate elements to pay attention to in a particular physics class. These different schemas will likely lead these two students to build connections between what they are learning in class and their other knowledge in different ways.

### 2) Epistemological resources

Resources exist not only at the knowledge level, but at the control level as well (Hammer and Elby, 2002). We make (often non-conscious) judgments about what knowledge to accept and what knowledge to call on to construct new knowledge based on partial information. When daddy comes home, his five year old daughter may tell him what's for dinner. If he asks, "How do you know?" he may get the response, "Mommy told me!" This is the epistemological resource, *knowledge as propagated stuff* – the idea that knowledge may be passed from one individual to another. If he asks, "What's your doll's name?" and gets the response, "Annie," the question, "How do you know?" may bring the response, "I made it up!"[1] This is *knowledge as fabricated stuff* – the idea that knowledge may be created. A wide variety of knowledge-building control resources are developed as one grows into adulthood.

## 3) Epistemic Games

Just as the choice of which resources to activate in a given situation can be controlled by schema based on partial and preliminary data, the choice of resources for building new knowledge can be controlled by epistemological schemas. When a student has a reasonably coherent schema for creating new knowledge (e.g., solving a problem) by using particular tools, we refer to it as an *epistemic game* or *e-game* (Collins and Ferguson, 1993; Tuminaro and Redish, 2007). These locally coherent (in time) cognitive activities typically:

- Have entry and ending conditions (How do you know when to play this game and how do you know when it's over?)
- Correspond to a particular knowledge base (What information will you call on?)
- Are guided by an external form or structure (Such as the writeup for a homework assignment – a derivation of a solution.)
- Have a restricted set of permitted moves (You can't pay attention to everything at once. The game picks a locally coherent set.)

Knowing e-games can be quite productive in solving problems if you recognize which games are appropriate. The research literature on problem solving (Hsu, Brewe, Foster and Harper, 2004) notes that one of the main problems novice problem solvers in physics have is in classifying problems appropriately – for example knowing whether to bring Newton's second law or energy conservation to bear (Chi, Feltovich and Glaser, 1989; Hardiman, Dufresne and Mestre, 1989; Sabella and Redish, 2007). Understanding what e-games students think are appropriate to play can help us create a nuanced understanding of what students know and what they need to know to become better problem solvers – more productive than simply saying they are "novice" or "expert."

The choice of e-game to play can be critical to students' success in problem solving. In one example (Tuminaro and Redish, 2007), a student was seen to play a game that is sometimes effective: *recursive-plug and chug*. In this game, one identifies a variable to be calculated, seeks an equation containing that variable, and sees if all the other symbols in that equation are given in the problem. If they are, one simply plugs into the equation and calculates the unknown. If there are additional unknowns, one seeks additional equations with those unknowns and goes through the process again until one has the same number of equations as unknowns. One then solves for all the unknowns mathematically. This e-game is often effective and is often taught explicitly in high schools.

Unfortunately, for slightly more subtle problems than simple numerical evaluation, some moves that are "missing" from this game may lead the student into deep trouble; moves such as "consider the mechanism of the phenomenon to see if the equation chosen is relevant," or "bring in quantitative information from one's personal experience." In an example cited in Tuminaro and Redish (2007), a student attempted to play recursive-plug-and-chug inappropriately. The problem asked her to estimate the difference in air pressure between the floor and ceiling of her dormitory room. In recursive-plug-and-chug, all information must come from an authoritative source – the textbook, lecture notes, or, best of all, the statement of the problem itself. The student chose the wrong equation ($PV=nRT$ rather than $P_d=P_0 + \rho g d$), didn't evaluate the mechanism, and sought to find the volume of her dorm room in the text of the problem. The only place a volume was mentioned was in the statement of the density of the air ("Take the density of air to be about 1 kg/m$^3$.") so she decided the volume of the dorm room had to be taken as 1 m$^3$. She made no effort to use her understanding of the concept of density to block this choice – or to use any of her conceptual knowledge (even about her dorm room) to evaluate her choices. The teaching assistant and the other students working with her on the prob-lem had a hard time talking her out of this. Her problem here was not a misunderstanding about the volume of the room, but rather an *epistemological misconception* – a misunderstanding of what knowledge to access and how to use it.

---

[1] Of course the child did not actually make up this name. He knew it from elsewhere. But the focus here is on the fact that the child made the choice of this name from all the names he knew. And the reason he knew this was the name is that he knew that he was free to make the choice and that was the choice he made.

Looking at Pellegrino's third principle in terms of mechanism makes us realize that thinking about thinking, choosing what to think about and evaluating one's thinking, is a complex process.

While we have laid out some of the components of a theory of mechanism in education, we want to be clear: While we advocate paying attention to mechanism, we do not mean to imply that thinking is mechanistic or can be modeled by a rigid production model. Individuals differ and differ from one time to another. The best we can hope for from a theory of education are probabilistic results. But we can infer some guidelines and hints for instructional approaches.

## 5. Implications for Educational Practice

This theoretical model, while supporting and giving mechanism for Pellegrino's three principles, has more detailed implications for thinking about learning, teaching, and the development of instruction. Here are six inferences that have important implications.

1. Students assemble their responses to instruction from what they already know – appropriately or inappropriately. This can lead to what appear to be preconceptions that are incorrect and robust. Note, however, that these may be created "on the fly" in response to new information that is being presented.
2. Although the strength of synapses can be adjusted somewhat, it is difficult to "undo" a strong synapse. Incorrect interpretations or associations can be made context dependent (suppressed in a particular context) or remapped but not easily removed. This together with 1. suggest that the removal of an inappropriate "preconception" may require some subtle instruction. (For an example, see Elby, 2001)
3. Student responses may be context dependent; students do not necessarily have broadly consistent (sometimes wrong) theories. The particular knowledge resources activated in a given circumstance can be dramatically affected by an individual's perception of contextual factors. This is known as context dependence. (This even implies that an individual may hold contradictory ideas or senses of a phenomenon without noting the contradiction.) Note that this can have important implications for assessment. "Split task" assessments, in which students are asked "mark the answer you think is wanted in this class but also mark (in a different way) the one you think makes most sense," often reveals surprising inconsistencies (McCasky, Dancy and Elby, 2004; Redish and Hammer, in preparation). (This is the way the Newton's third law evaluation discussed above was made.)
4. In order to design effective instruction, it does not suffice to know what students know about the topic; one must know how they associate elements of their knowledge and how they control access to it through their expectations.
5. In order to understand student difficulties, expert instructors may have to "reverse engineer" their compiled knowledge.

Students' framing as to what knowledge is appropriate for a given task is often automatic and unnoticed by the student. This can result in their failure to access knowledge that they have that is appropriate or even critical. Identifying such "epistemological preconceptions" may be critical in designing instruction to develop broader learning skills – helping students to define learning goals and monitor and evaluate their progress.

## 6. Thinking and Learning about Modeling

### A. *An example – use of math in science and engineering.*

All the issues discussed in section V come into play when we think about developing a general skill. For this paper, we use *learning to use math effectively* as one example of skill development in science and engineering. To do this, students not only have to acquire particular mathematical knowl-edge, they have to learn to use it appropriately and effectively in a scientific context. This is an essential component of developing adaptive expertise in engineering.

This is subtler than it seems to an expert who has already mastered that skill. To understand where the students have to get to, we need to understand how mathematics is embedded in the broader web of scientific knowledge and we need to unpack elements of the experts' compiled knowledge that may have become invisible to them. A critical element is that expert modelers use their physical knowledge of the system in choosing, thinking about, solving, and interpreting the mathematics they use. We begin by looking at the differences between abstract math and math in a physical context.

---

**The Montillation of Traxoline**

It is very important that you learn about traxoline. Traxoline is a new form of zionter. It is montilled in Ceristanna. The Ceristannians gristerate large amounts of fevon and then bracter it to quasel traxoline. Traxoline may well be one of our most lukized snezlaus in the future because of our zionter lescelidge.

*Answer the following questions in complete sentences. Be sure to use your best handwriting.*

—1. What is a traxoline?

—2. Where is traxoline montilled?

—3. How is traxoline quaselled?

—4. Why is it important to know about traxoline?

---

Figure 3. A sample standardized science test problem. *Attributed to Judith Lanier*

This is often used as an example of a "bad" problem as almost anyone who is competent in English can solve these problems without having any idea what the terms mean. But this judgment neglects the important point that a significant amount of grammatical and syntactic knowledge is being tested here. You could not solve these problems if you did not understand the difference between subjects and objects, how to identify verb tenses and endings, the role and implication of helping verbs, etc., etc., etc.

There is a strong analogy between this example and the use of mathematics in science. The mathematics in itself is about grammatical and syntactic relationships – and permits the drawing of complex conclusions about the placeholders (variables) without having any idea what those placeholders stand for. This example sends a strong message to scientists and engineers: We not only expect our students to be able to understand mathematics (syntax); we expect them to combine this knowledge with knowledge of what the math is talking about in a tightly integrated way – on the *meaning* of the symbols. This is different from straight math and can even lead to differences in the way equations are interpreted.

**B. *The difference between meaning in physics and math: A shibboleth[2]***

Try the following problem on your friends and colleagues in engineering, math, and physics (Dray and Manogue, 2002).

---

One of your colleagues is measuring the temperature of a plate of metal placed above an outlet pipe that emits cool air. The result can be well described in Cartesian coordinates by the function

$$T(x, y) = k\left(x^2 + y^2\right)$$

where $k$ is a constant. If you were asked to give the following function, what would you write?

$$T(r, \theta) = ?$$

---

Figure 4. A problem whose answer tends to distinguish engineers and physicists from mathematicians.

---

The amusing thing about this problem is that you are very likely to get different answers from different populations. An engineer or physicist who works regularly with polar coordinates is likely to give the response $T(r,\theta) = kr^2$, a result obtained by assuming the variables are related by the familiar polar to Cartesian relation $r^2 = x^2 + y^2$. A more mathematical colleague is likely to respond, $T(r,\theta) = k\left(r^2 + \theta^2\right)$. The function is defined mathematically by saying, "Add the squares of the two variables and multiply by $k$." The engineer will object. "You can't add $r^2$ and $\theta^2$! They have different units." The mathematician might reply, "No problem. I see what you mean. You just have to change the name so that each symbol represents a unique functional dependence. For example, you could write $T(x,y) = S(r,\theta) = kr^2$." Unfortunately, your more concrete friend (imagine a chemical engineer at this point) is still unlikely to be satisfied. "You can't write the temperature equals the entropy! That will confuse everything."

Many engineers, physicists, and mathematicians are surprised by this story. Each side believes that its way of using equations is the "obvious way" and that surely even a mathematician (an engineer) would agree. Unfortunately, that is not the case. Each group strongly prefers its own interpretation of how to write an equation.

These two examples dramatically illustrate that fact that in science and engineering, we tend to look at mathematics in a different way from the way mathematicians do. The mental resources that are associated (and even compiled) by the two groups are dramatically different. The epistemic games we want our students to choose in using math in science requires the blending of distinct local coherences: our understanding of the rules of mathematics and our sense and intuitions of the physical world.

Here are some of the ways the use of math differs in science and engineering from the way it is often taught in a math class.

- Equations represent relationships among physical variables, which are often empirical measurements.
- Symbols in equations carry information about the nature of the measurement beyond simply its value. This information may affect the way the equation is interpreted and used.
- Functions in science and engineering tend to stand for relations among physical variables, independent of the way those variables are represented.

The last of these is responsible for the varied responses to the shibboleth. The engineer or scientist views the function as telling how warm or cool the plate is at a particular point in space. The resolution of that function into numbers can take place in a variety of ways depending on the particular coordinate system chosen to represent the physical reality.[3] Professional scientists and engineers typically interpret what appear on the surface to be mathematical equations as physical. There is a strong association between their mental spaces that conceptualize the physical system and the mathematical symbols that represent various measurements. Students often fail to make this connection.

## C. *A model of mathematical modeling*

If we want our students to learn not just the skill of mathematical manipulation but also the skill of using math in science effectively, we have to unpack our compiled knowledge and develop a deeper understanding of mathematical modeling, reverse engineering our expertise. One way of describing the often-implicit processes that are used by scientists and engineers in creating, applying, and evaluating mathematical models is shown schematically in the model shown in Figure 5.

---

[3] Of course there are situations in which scientists and engineers pay particular attention to the mathematical dependence of a function. Notable are the switches in mechanics from Lagrangian to Hamiltonian and in physical chemistry between different chemical potentials.

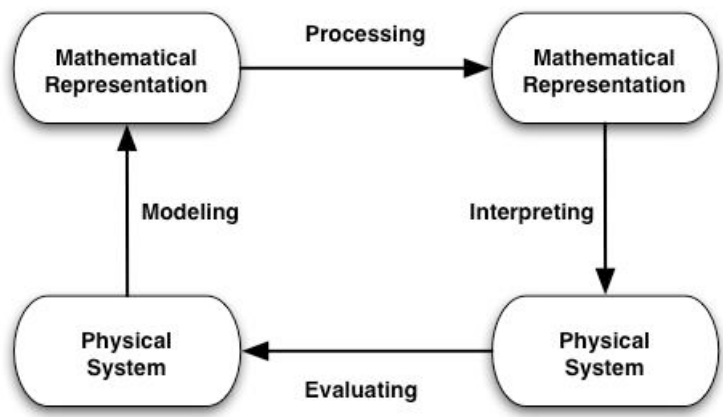

Figure 5. A model for modeling a physical system with mathematics.

We begin in the lower left corner by choosing a physical system we want to describe. Within this box, we have to decide what characteristics of the system to pay attention to and what to ignore. This is a crucial step and is where much of the skill or "art" in using math in STEM lies. Looking at a complex physical system and deciding what are the critical elements that must be kept, what are trivial effects that can be ignored, and what are somewhat important effects that can be ignored at first and corrected later is what we mean by "getting the physics right." Einstein said it best: "*Everything should be as simple as possible – but not simpler*."

Once we have decided what we need to consider, we then *model*. We map our physical structures into mathematical ones – we create a mathematical model. To do this, we have to understand what mathematical structures are available and what aspects of them are relevant to the physical characteristics we are trying to model.

Now that we have mathematized our system, we are ready to *process*. We can use the technology associated with the math structures we have chosen to transform our initial description. We may be solving an equation or deriving new ones. But once we have done that, we are a long way from finished. We still have to *interpret*. We see what our results tell us about our system in physical terms and then *evaluate*. We have to decide whether our results adequately describe our physical system or whether we have to modify our model. Note that this model should not be interpreted as a flow chart. There are cross-links throughout. For example, because of the fact that the equations are physical rather than purely mathematical, the processing can be affected by physical interpretations.

### D. *The games we teach, the games they play*

Three examples of common practice, when viewed in terms of the theoretical structures described here, are seen to be possibly counterproductive to our newly stated goals of developing general skills and adaptive expertise. They are: focusing on algorithms and results, substituting numbers in immediately, and concentrating on the processing leg of the modeling diagram in homework and exams.

Often, in both engineering and physics classes, we tend to focus our instruction on process and results. When we teach algorithms without derivation, we send our students the message that "only the rule matters" and that the connection between the equation we use in practice and the assumptions and scientific principles that are responsible for the rule are irrelevant. Such practices may help students produce results quickly and efficiently, but at the cost of developing general and productive associations and epistemic games that help them know how their new knowledge relates to other things they know and when to use it. As narrow games get locked in and tied to particular contexts, students lose the opportunity to develop the flexibility and the general skills needing to develop "adaptive expertise."

A similar thing happens when students are permitted (sometimes encouraged) to put numbers in immediately before solving a problem in order to "make the math simpler." Problems indeed become more math-like but they lose the connection to the underlying physical model. If we are to build modeling skills, the emphasis needs to be on the relationships among variables and parameters. These are lost when input variables and parameters are combined numerically. This approach not only loses opportunities for the students to build cognitive links between the physical world and the math, it limits creativity and the development of design skills. If the result desired is not what we want, it is much harder to see how it might be improved by a modified design if the roles played by the adjustable parameters are not visible in the expression for the result.

A third problem occurs when the processing leg of the modeling diagram (Figure 5) is emphasized and the others are essentially ignored. Students become apparently fluent in the techniques emphasized in the class, can do well in the homework and exams, but because they lack robust links to a broad knowledge-structure based firmly in the physical world, their knowledge is strongly context dependent and appears brittle. When they move on to a new situation (as, for example, in another class), they are unable to call on the knowledge they appeared to learn effectively in the previous class. The instructor may find it hard to realize what has happened if his or her own process knowledge is strongly bound to physical knowledge. The instructor makes the appropriate connections and interpretations to the work the student presents but may not ask rich enough questions to reveal that the student has not made the same connections.

These issues are not new. Many teachers are aware of them. Our theoretical analysis provides a way of understanding their insights, connecting them to the growing knowledge of how people think and learn, and communicating them to others. They have been addressed by teachers in a variety of ways. Two approaches used by the authors (and others) are

- by means of a course explicitly devoted to modeling and epistemological development (Redish and Hammer, in preparation; Smith and Starfield, 1993).
- by diversifying the types of problems given in the context of a more traditional class (Koen, 2003; Redish, 2008).

We discuss these approaches in the next section.

## 7. Implications for Instruction and Curriculum Design

### A. *Building student's modeling skills*

We have only discussed a part of the emerging theoretical structure for education here: the part associated with the thinking of the individual. But addressing the cognitive issues we want to impact in practice requires that we also develop a theoretical understanding of the classroom and how the individual interacts with and responds to the instructional environment. Extensive research supports the idea that students can develop their knowledge and skills much more effectively in well-structured group-learning environments than they can individually (Johnson, Johnson and Smith, 1991; Smith, Sheppard, Johnson and Johnson, 2005; Johnson, Johnson and Smith, 2007). The combination of an approach to modeling that focuses on linking mathematical and physical knowledge structures with a carefully designed group-learning environment was described by Smith and Starfield (Smith and Starfield, 1993; Starfield, Smith and Bleloch, 1990).

First year engineering students in Starfield & Smith's 100 or so student *How to Model It* course are organized into groups of three (using a stratified random assignment approach based on familiarity with the use of spreadsheets) within the first 10 minutes of their first class. This arrangement surprises some of the students. However, most of them eagerly find the other members of their group and introduce themselves. A task that requires them to consider and quantify a physical system and tie their personal experience to mathematics (such as estimating the maximum number of ping-pong balls that could fit in the room) is assigned and one copy is given to each group.[4] A transparency and pen is given to each group so they can prepare their formulation for presentation to the class. Instructions to the students include: each group is to formulate one answer to present, make sure everyone participates, and make sure everyone can explain your group's answer, especially the model used, the assumptions and the mathematical formulation.

---

[4] This problem was introduced to the engineering education community by Billy Koen in about 1981. See also (Koen, 2003).

While the students are working on the problem, we circulate among the groups; listening as the students discuss the problem with each other, occasionally intervening to ask a student to explain, and continually providing support and encouragement for the group work.

After we call the whole group back together, individual students are randomly selected to give their group's answer and explain the method used to arrive at their answer. Several answers are requested and recorded on an overhead. Answers and methods are compared in terms of the formulation and assumptions. The usefulness of formulating an algorithm and using a notation system are discussed. Also, we engage the students in reflection about (1) what features their model captures (and leaves out) (2) how good an answer is needed, i.e., what is the purpose?, and (3) what they would do if they needed a better answer. Since most groups come up with a single-point estimate, we also ask them to consider the heuristic of establishing lower and upper bounds for their answer.

The final step in the cooperative groups involves the students processing their work in two ways. First they discuss how well they solved the problem, including their use of strategy and how confident each member feels about the group's answer and method. Second, they process how well they worked as a group – what things went well and what things they need to work on to function more effectively together.

The essence of this classroom structure is students working together to get the job done. The book, *Active Learning: Cooperation in the College Classroom*, carefully defines cooperative learning, describes how to do it, and provides detailed rationale (Johnson, Johnson and Smith, 2006).

In the *How to Model It* class and more broadly it is argued that modeling and problem-solving are inseparable, so much so that it is difficult to learn to solve problems without learning to model, and vice versa. We take the view that modeling is a more specific goal, that it is easier to learn to model, that it is useful to learn to model, and that incidentally one also learns a lot about problem-solving. Most problems are "solved" by constructing a representation (a mathematical expression, a graph, a manual or computer simulation program, a physical model, etc.). The process can be roughly described as following the steps of Fig. 5: (1) Starting with a real system and formulating a model, (2) Drawing conclusions from the model using reasoning and mathematical processing, (3) Interpreting the model conclusions, and (4) Evaluating and validating that the model actually works.

Modeling means constructing a simplified representation of some physical, biological or social phenomenon that is too complicated to represent in all the details of its entirety. Students manipulate their model to gain understanding of the phenomenon being modeled. Students are offered advice along the way, including heuristics such as:

1. Keep it simple. Beware of building a complicated model when a simple one will do.
2. Try to understand the problem. Imagine that you're in the situation--what does it look and feel like? how are things changing?
3. The principal benefit of modeling is often associated with what is learned while trying to build the model.
4. GIGO: Garbage-In-Garbage-Out. A model is not any better than the information that goes into it.

In this course, it is stressed that learning in all disciplines involves constructing models, investigating ideas and developing problem-solving skills. These activities are not limited to students in science and engineering. They are shared by all who have a desire to understand, to interpret, and to explain.

Recent work in this area has focused on developing tasks to engage students in the modeling process (Diefes-Dux, Moore, Zawojewski, Imbrie and Follman, 2004; Moore and Diefes-Dux, 2004). The tasks are described as Model-Eliciting Activities (MEAs) and a set of principles for developing them is available in Lesh and Dorr (2003).

### B. *Diversifying problems in a more traditional course*

Our theoretical structure emphasizes that knowledge naturally develops context dependence. If we want the modeling skills developed in a course such as described above to be a general part of a student's repertoire, they have to be called on and reinforced in a variety of contexts. This means that the skills of quantifying one's personal experience, of making mathematical models of physical systems, and of interpreting mathematical equations and results in terms of physical reality must be called upon in other contexts. Below, we offer two examples (drawn from the physics problem collections of one of us (Redish, 2008)).

Often, we assume idealizations that simplify the mathematics, allowing us to create straightforward (and possibly unrealistic) solutions. For example, in most physics classes, springs are treated as ideal Hooke's law oscillators. The problem shown in Figure 6 asks the student to consider how a real spring would behave, thereby encouraging them to contemplate what aspects of physical reality is being modeled by Hooke's law.

When we consider the properties of a spring, we typically imagine an ideal spring, which perfectly satisfies Hooke's law, T = kΔs. In actual fact, this is an awful approximation for most extensions or compressions of the spring. What we typically do is stretch the spring some amount beyond its rest length, say by hanging a weight from it. For small displacements around that equilibrium position, the extra force exerted by the spring is linear around the resting point: "F = -kx". 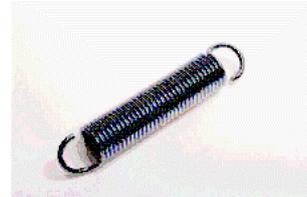

Consider the spring shown in the figure. Its resting length is 5 cm. Consider that a pair of equal and opposite forces of magnitude, T, is exerted on the two ends of the spring, pulling or pushing in opposite directions. If T is positive, it means the forces are pulling to try to stretch the spring. If T is negative, it means they are trying to compress the spring.

Sketch a graph of how the length of the spring varies as a function of T, considering both positive and negative values of T and going to very large values. Make plausible guesses for the values of the length when unusual things happen. Identify the "Hooke's law" regimes and identify salient features of your graph, described what is happening at those salient points.

Figure 6.

A problem that goes the "other way" – asking the student to interpret the structure of a mathematical in physical reality is given in Fig. 7.

The pair of coupled ordinary differential equations

$$\frac{dx}{dt} = Ax - Bxy$$

$$\frac{dy}{dt} = -Cy + Dxy$$

is referred to as the Lotka equation and is supposed to represent the evolution of the populations of a predator and its prey as a function of time.
(a) Which of the variables, *x* or *y*, represents the predator? Which represents the prey? What reasons do you have for your choice?
(b) What do the parameters *A, B, C,* and *D* represent? Why do you say so?
(c) How good a model do you think this is? Can you identify some phenomena that might be important that are not included in this mathematical model?

Figure 7.

Problems of these types, calling on other legs of our modeling model than simply processing, could be helpful in beginning to move even our traditional courses towards helping students to more skill-oriented learning that is both robust and agile.

## 8. Conclusions

Reconsideration of the environments in which today's engineering students will work, leads to new goals for their education (National Academy of Engineering, 2004, 2005). To achieve these goals, Pellegrino (2006) has called for a central theory of the nature of learning in a given domain of knowledge to provide an alignment of the critical components of curriculum, instruction, and assessment.

In this paper, we have outlined some of the components of the growing theoretical understanding of the cognitive process being developed in neural, cognitive, the behavioral, and educational sciences. Our claim is that a theoretical framework of thinking and learning based on organization of cognitive resources allows us to make progress towards such a central theory for engineering education. To demonstrate how the general framework of cognitive resources using the component concepts of activation, association, compilation, and control can be applied in the context of a particular knowledge domain, we have considered the example of blending two knowledge structures – mathematics and physical knowledge – into the engineering skill of modeling. Analyzing this skill in terms of our theoretical framework gives new ways of analysis that permits the identification of specific issues and difficulties that have important educational implications. We review briefly a few instructional methods that the authors have used in order to address the development of this skill.

Much more needs to be done, including the use of the framework to develop new assessments and its expansion to integrate the understandings that have been developed as to how instructional environments function and interact with cognitive development. But we believe that the core elements of a theoretical framework based on cognitive science are in place and can serve as the basis for a more rigorous and effective process of instructional reform.

## Acknowledgements

This material is based upon work supported by the US National Science Foundation under Awards No. DUE-05-24987 and REC-04-40113. Any opinions, findings, and conclusions or recommendations expressed in this publication are those of the author(s) and do not necessarily reflect the views of the National Science Foundation.

**Author Biographies**

Edward F. Redish is Professor of Physics and an Affiliate Professor in Curriculum & Instruction at the University of Maryland, College Park. He has been involved in research and development in physics education for over 20 years. He is a fellow of the American Physical Society and the AAAS. He has received many awards for his work in education including the Robert A. Millikan Medal from the AAPT and the NSF Director's award as a Distinguished Teaching Scholar.. Currently he is doing research on student epistemologies and expectations, on cognitive models of student thinking in physics, and on student difficulties with the use of mathematics in physics, especially upper division physics

Karl A. Smith, Cooperative Learning Professor of Engineering Education, Department of Engineering Education, and Fellow, Discovery Learning Center at Purdue University West Lafayette. He has been at the University of Minnesota since 1972 and is in phased retirement as Morse-Alumni Distinguished Professor of Civil Engineering. Karl has worked with thousands of faculty all over the world on pedagogies of engagement, especially cooperative learning, problem-based learning, and constructive controversy. He has co-written eight books including *Cooperative learning: Increasing college faculty instructional productivity*; *Strategies for energizing large classes: From small groups to learning communities*; and *Teamwork and project management, 3$^{rd}$ Ed.*